\pdfoutput=1
\documentclass[conference]{IEEEtran}
\usepackage{booktabs}
\usepackage{multirow}
\usepackage{graphicx}
\usepackage{enumitem,kantlipsum}
\usepackage{subfigure}
\usepackage{grffile} % fix figure file extention issue
\usepackage{color}
\usepackage{xcolor}
\usepackage{xspace}
\newcommand{\myfigure}[1]{Figure~#1}
\newcommand{\mytable}[1]{Table~#1}

\newcommand{\ie}{\textit{i}.\textit{e}., }
\newcommand{\eg}{\textit{e}.\textit{g}., }

\newcommand{\scout}{{\sc Scout}\xspace}
\newcommand{\micky}{{\sc Micky}\xspace}
\usepackage{url}

\begin{document}
\title{
Micky: A Cheaper Alternative for Selecting Cloud Instances}

\author{\IEEEauthorblockN{Chin-Jung Hsu\textsuperscript{*}, 
Vivek Nair\textsuperscript{*}, 
Tim Menzies,
and Vincent Freeh}
\IEEEauthorblockA{Department of Computer Science\\
North Carolina State University\\
Raleigh, North Carolina\\
Email: \{chsu6, vnair2, tjmenzie, vwfreeh\}@ncsu.edu}
}


%Telephone: (800) 555--1212, Fax: (888) 555--1212}


\maketitle

\begin{abstract}

Most cloud computing optimizers explore and improve one workload at a time.  When optimizing many workloads, the \emph{single-optimizer} approach can be prohibitively expensive. Accordingly, we examine ``collective optimizer'' that concurrently explore and improve a set of workloads significantly reducing the measurement costs.
Our large-scale empirical study shows that there is often
a single cloud configuration which is surprisingly near-optimal for most workloads.
Consequently, we create a \emph{collective-optimizer}, \micky, that reformulates the task of finding the near-optimal cloud configuration as a multi-armed bandit problem. MICKY efficiently balances exploration (of new cloud configurations) and exploitation (of known good cloud configuration).
Our experiments show that MICKY can achieve on average 8.6 times reduction in measurement cost as compared to the state-of-the-art method while finding near-optimal solutions.

Hence we propose \micky as the basis of a practical collective optimization method for finding good cloud configurations (based on various constraints such as budget and
tolerance to near-optimal configurations).

\end{abstract}


% \fi
\IEEEpeerreviewmaketitle

\newcommand\blfootnote[1]{%
  \begingroup
  \renewcommand\thefootnote{}\footnote{#1}%
  \addtocounter{footnote}{-1}%
  \endgroup
}
\blfootnote{\textsuperscript{*}These two authors contribute equally to the work.}

%%%%%%%%%%%%%%%%%%%%%%%%%%%%%%%%%%%%%%%%%%%%%%%%%%%%%%%%%%%%%%%%%%%%%%%%
%  Introduction
%%%%%%%%%%%%%%%%%%%%%%%%%%%%%%%%%%%%%%%%%%%%%%%%%%%%%%%%%%%%%%%%%%%%%%%%

\section{Introduction}

\iffalse
\textcolor{red}{
TODO:
\begin{itemize}
    \item we only talk about deployment cost but we do not talk about execution time. We need to mention why we do this
    \item need to make sure we are using the same terminology. I use suitable cloud configuration, deployment cost, search cost, and execution time.
    \item we need to make sure we explicitly mention what is VM Type, what is cloud instance, what is cloud configuration, what is workload.
    \item CherryPick and \micky should have the same color lines---in Figure 4 and 5
\end{itemize}
}
\fi

%Define Cloud computing optimizer xxx
Cloud computing optimizer is a device to select the best cloud configurations
(such as virtual machine (VM) types and the number of VMs)
for a given workload.
Choosing the right cloud configuration is essential to maximize application performance and minimize operational costs. However, such optimization task is not straightforward due to opaque resource requirement~\cite{Yadwadkar2017,Hsu2016}.
To address this challenge, prior work either
builds prediction models (as in \emph{Ernest}~\cite{Venkataraman2016} and \emph{PARIS}~\cite{Yadwadkar2017}) or
uses sequential model-based optimization (as in \emph{CherryPick}~\cite{Alipourfard2017}, \emph{Arrow}~\cite{Hsu2017} and \emph{Scout}~\cite{Hsu2018scout}).

While they are effective, they are only designed for a single workload.
In practice, it is rare to migrate only one workload~\cite{khajeh2010cloud,sripanidkulchai2010clouds}.
Since these optimizers are expensive to run,
applying them independently to workloads requires
significant measurement cost and
long optimization process.
In this paper, we optimize a batch of workloads altogether.

This kind of collective optimization is impossible if
workloads execute very differently on different cloud configurations.
Prior work reports there does not exist an one-size-fits-all VM type that is best for all workloads~\cite{Alipourfard2017, Yadwadkar2017, Hsu2017}.
However, while analyzing the data from our large empirical study involving three different software systems and over 100 workloads, we noticed that there does exist at least a cloud configuration (\eg{m4.large}), which performs satisfactorily
for the majority of workloads.
If the above is prevalent in cloud computing,
it should be possible to simplify collective optimization.
In this paper, we exploit this phenomenon in order to further reduce optimization cost.

We call such a cloud configuration \textit{Exemplar Configuration}, which is near-optimal or satisfactory for the majority of workloads.
In our empirical study, the exemplar configuration is only 5-20\% slower or more expensive than the optimal choice.
In any cloud optimizer, there exists a trade-off between search performance (how far a choice is from the optimal) and measurement cost (how many tests an optimizer requires to find a suitable configuration).
With the exemplar configuration,
we can trade a slight decrease in search performance for a large reduction in measurement cost because
redundant efforts can be reduced in collective optimization.
When optimizing a group of workloads,
such trade-off not only brings significant cost reduction but also shortens the optimization process as well as the migration procedure.
However, finding such an exemplar configuration is not straightforward because it depends on workloads and performance objectives.
Moreover, as cloud providers expand their cloud portfolio, the exemplar configuration is also likely to change.
In this paper, we focus on finding out this exemplar configuration efficiently.

To this end, we propose and evaluate a collective optimization method, \micky\footnote{Micky (Rosa) is a  character, from the Hollywood movie 21, who founded the MIT Black Jack team of card counters.}, which enables users to deploy a group (not one) of workloads to the cloud more efficiently (lower measurement cost).
We reformulate ``finding the exemplar configuration'' as the multi-armed bandit problem~\cite{weber1992gittins,bergemann2006bandit,audibert2011introduction,dambreville2017load,jiang2017pytheas}.
The two problems are similar because
the bandit problem aims to maximize rewards (\micky, for example, minimizes execution time or operational cost) in a series of decisions (to run a workload on a cloud configuration), each is associated with an unknown payoff and a known opportunity loss (whether the decision meets the performance objective).
Our evaluation shows that \micky can find the exemplar configuration using only 12\% of the total effort compared to a sophisticated single-optimizer.
This cooperative style of search methods ensures that users do not need to optimize each workload separately; instead, finds the exemplar cloud configuration collectively, thereby reducing measurement cost. 

\micky finds a configuration that is near-optimal for the majority of workloads.
But the chosen configuration could perform unacceptably for some workloads.
To remedy this issue,
we integrate our previously built system \scout
to identify sub-optimal cases~\cite{Hsu2018scout}.
This enables elaborate optimization for unsatisfactory workloads if strict performance is required.

We demonstrate the effectiveness of \micky by evaluating it on 107 real-world workloads (using three popular software systems) and show that \micky can find near-optimal cloud configurations by using only a fraction (12\%) of the measurement cost used by the state of the art methods, at the expense of less optimal choices.
There is always a trade-off between search performance and measurement cost.
Based on our evaluation,
we advise users not to use \micky
only when the same workloads will repeat  more than tens of runs (\ie{30 times} using our analysis) .
To deploy a batch of workloads to cloud,
we believe \micky is more desirable than state-of-the-art methods
because the higher number of recurrence would certainly
limit the applicability of cloud optimization.
Furthermore, those sub-optimal choices can be eliminated
through the integration between \micky and \scout,
thereby creating a more robust solution.

The main contributions of this paper are:

\begin{itemize}
    \item Using a large-scale empirical study, we discover the exemplar cloud configuration (Section~\ref{sec:empirical}).
    \item We propose to formulate ``finding the exemplar configuration in the cloud'' as the multi-armed bandit problem (Section~\ref{sec:formulation}).
    \item Using real-world data collected from EC2, we show \micky can quickly find the exemplar configuration (Section~\ref{sec:evaluation}).
    \item We show the system integration between \micky and \scout is able to achieve low optimization cost and high performance guarantee (Section~\ref{sec:system}).
     \item We design a practical guide for selecting the right cloud computing optimizer~(Section~\ref{sec:guide}).

\end{itemize}

%%%%%%%%%%%%%%%%%%%%%%%%%%%%%%%%%%%%%%%%%%%%%%%%%%%%%%%%%%%%%%%%%%%%%%%%
%  Motivation
%%%%%%%%%%%%%%%%%%%%%%%%%%%%%%%%%%%%%%%%%%%%%%%%%%%%%%%%%%%%%%%%%%%%%%%%

\section{Why Collective Optimization}
\label{sec:motivation}

A cloud optimizer is often evaluated with search performance and measurement cost.

\textbf{Search performance}
is the measure of the quality of the found solutions by an optimizer.
For example, in searching for the most cost-effective configuration, an optimizer that finds a configuration that is only 10\% more expensive than the optimal is considered better than another optimizer
that can only find a configuration that is 30\% slower.
In this paper, we use normalized performance (to the optimal) for evaluation.

\textbf{Measurement cost}
is the total cost of running an optimizer.
An optimization process is expensive because it requires
to test a workload on some cloud configurations for deriving
the best choice.
We use the number of tests as the measurement cost
because it is an intuitive measure.
The amount of charge is another measure~\cite{Alipourfard2017}.

There is always a trade-off between measurement cost and search performance.
The primary motivation for collective optimization is to reduce high measurement cost of optimizing multiple workloads.
If users demand strict search performance, they better turn to single-optimizers.
However, we argue that collective optimization is promising
because it achieves comparable or slightly worse search performance
while reducing measurement cost significantly.
In the following,
we discuss the benefits of having a collective optimizer.

\textbf{Large scale cloud migration.}
Cloud computing is a cost-effective solution.
Enterprises are moving in-house applications to the cloud,
and need a quick way for large migration~\cite{khajeh2010cloud,sripanidkulchai2010clouds}.
Elaborate optimizers are expensive (in measurement cost) and time-consuming (in optimization process).

\textbf{Limited budgets.}
The single-optimizer such as \emph{CherryPick} and \emph{Scout} are effective and desirable for highly recurring workloads because the measurement cost can be amortized.
However, the number of budgets to run optimizers
does not increase linearly with the number of workloads.
To better support multiple workloads,
we need to reduce measurement cost while delivering comparable search performance.

\textbf{Expanding cloud portfolio.}
Cloud providers expand their cloud portfolio more than 20 times in a year~\cite{ec2history}.
Therefore, users have to rerun optimizers to update their configurations for all workloads.
Again, this is an expensive and time-consuming process.

\textbf{Seed cloud optimizers.}
All the cloud optimizers require initial measurements.
It is unclear how to determine the best starting points.
In this paper, we aim to find the exemplar configurations,
which can be used as the starting points, thereby
reducing measurement cost.
The exemplar configuration can be used to seed singe-optimizers such as CherryPick and \scout, which will be discussed more in Section~\ref{sec:system}.

In summary, users would prefer collective optimization if search performance is comparable to single-optimizers while measurement cost can be reduced greatly.

%%%%%%%%%%%%%%%%%%%%%%%%%%%%%%%%%%%%%%%%%%%%%%%%%%%%%%%%%%%%%%%%%%%%%%%%
%  MAB
%%%%%%%%%%%%%%%%%%%%%%%%%%%%%%%%%%%%%%%%%%%%%%%%%%%%%%%%%%%%%%%%%%%%%%%%

\begin{figure*}[t]
 \centering
 \subfigure[Hadoop 2.7]{
 \label{fig:motivation_percentage_1}
 \includegraphics[width=.3\textwidth]{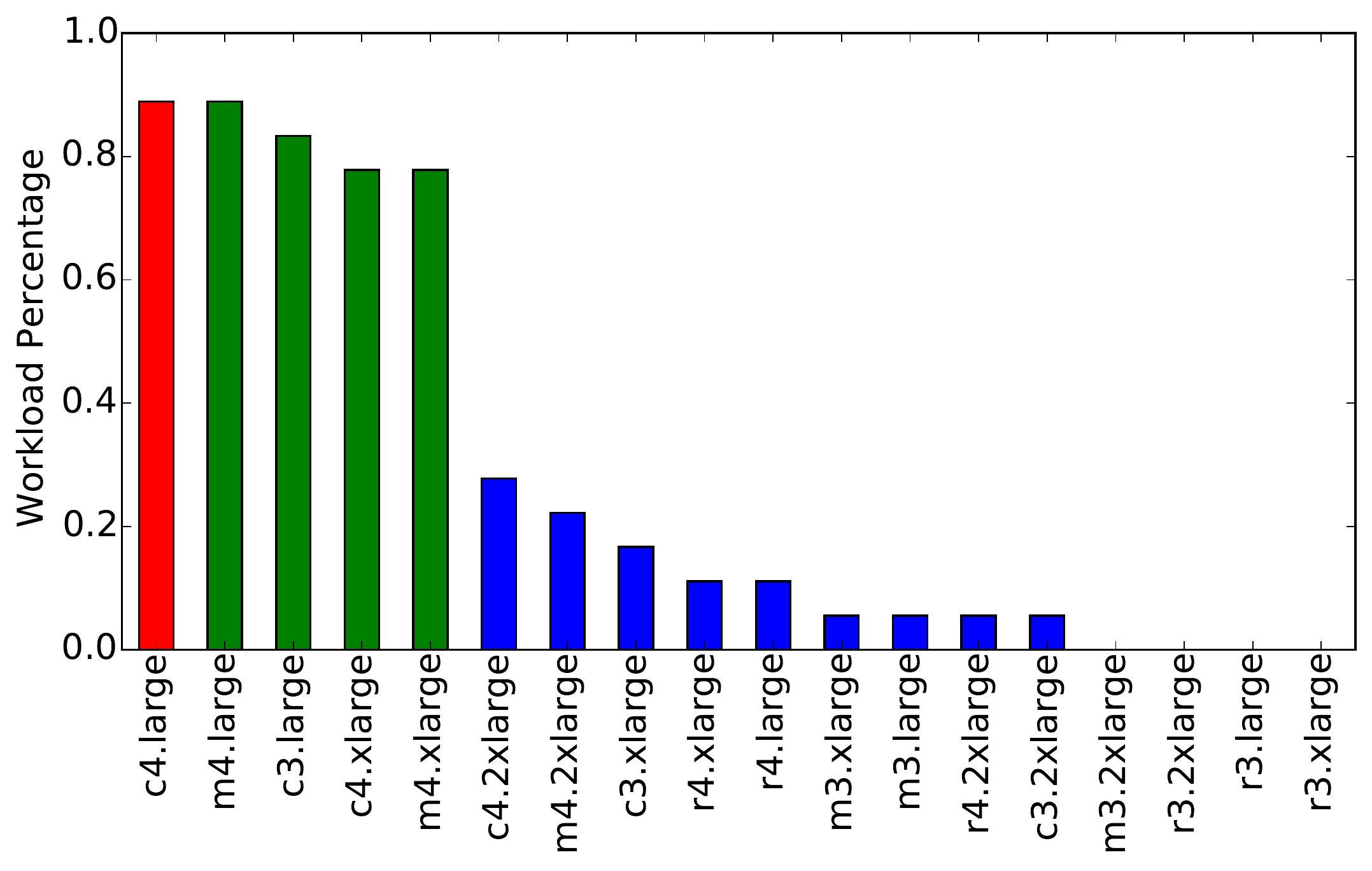}
 }
 %\hspace{.1\textwidth}
 \subfigure[Spark 2.1]{
 \label{fig:motivation_percentage_2}
 \includegraphics[width=.3\textwidth]{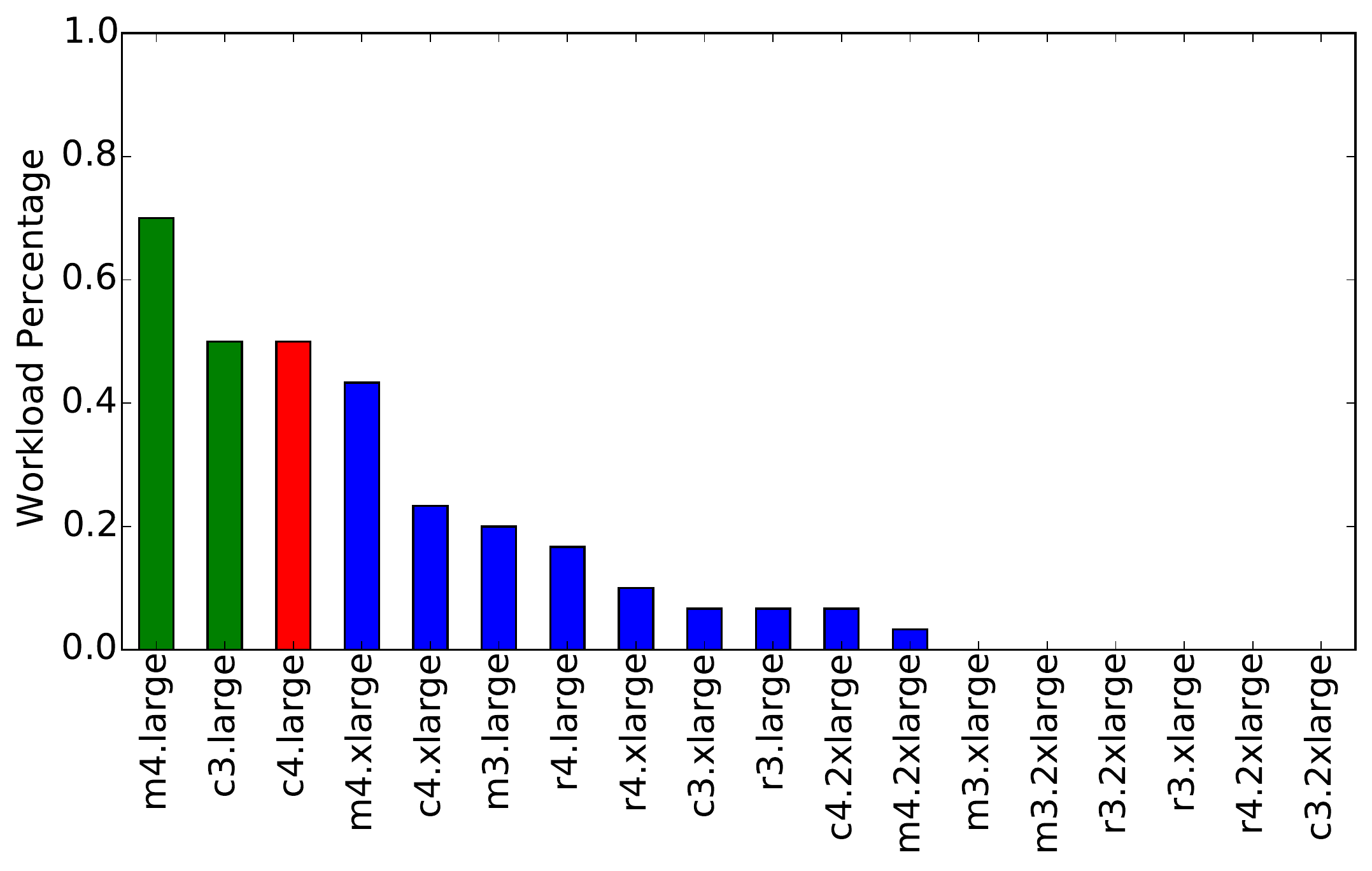}
 }
 \subfigure[Spark 1.5]{
 \label{fig:motivation_percentage_3}
 \includegraphics[width=.3\textwidth]{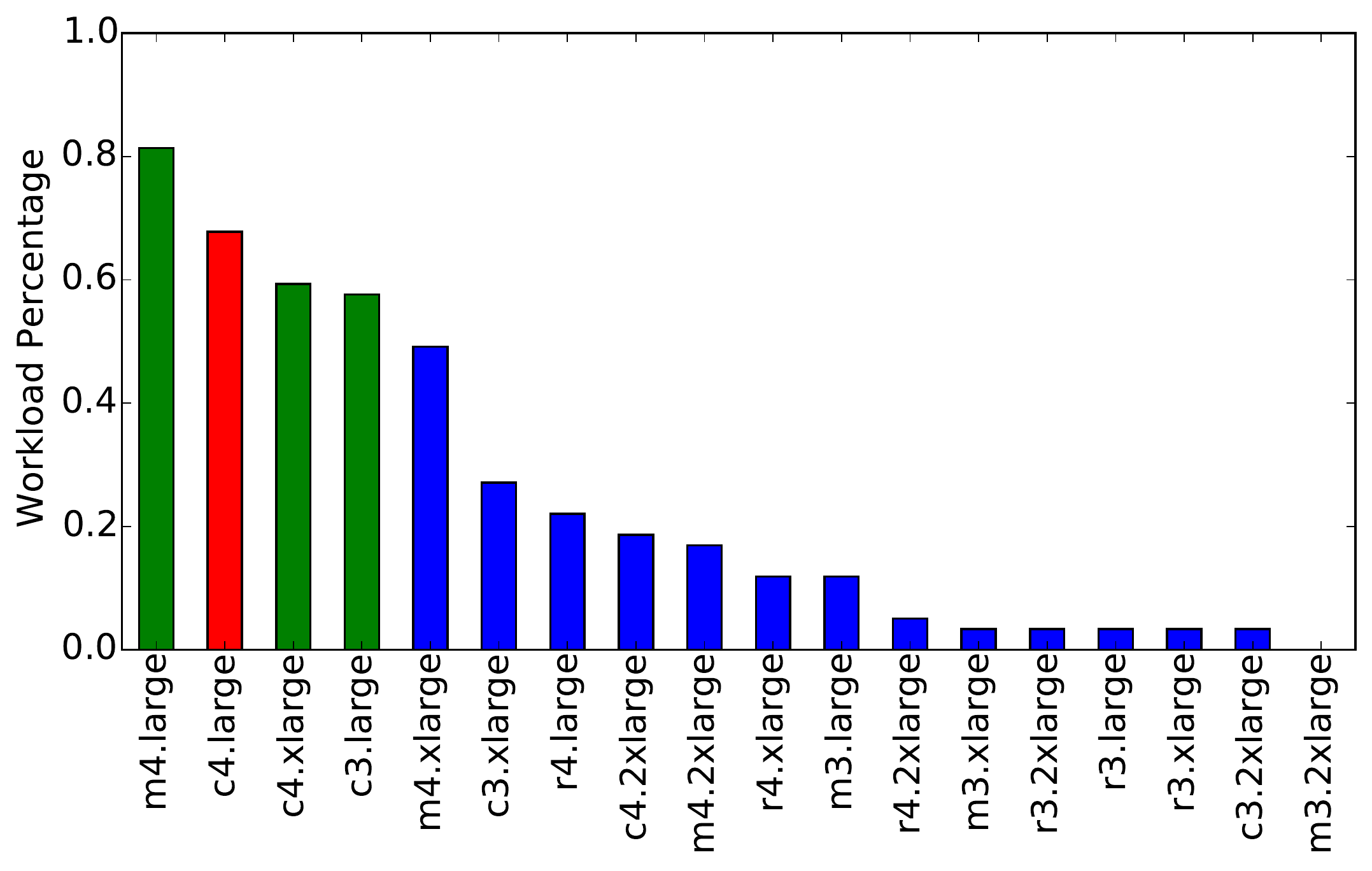}
 }
 \vspace*{-2mm}
 \centering
 \caption{\textbf{Opportunity to find the exemplar VM instances across workloads for reducing operational cost.} The \emph{y-axis} represents the percentage of workloads (out of 107 in three systems) that are within 30\% difference with the optimal performance.  The colored bars are VM types that considered the exemplar configurations for the majority of workloads ($>= 50\%$).
 The \textcolor{red}{red} bar represents that the VM type is more likely to be the optimal choice.}
 \label{fig:motivation_percentage}
 \vspace*{-4mm}
\end{figure*}

\section{Finding the Exemplar Cloud Configuration}
\label{sec:mab}

In this section, we first present our empirical study
on investigating the potential of finding the exemplar cloud configuration.
We then formulate ``finding the exemplar configuration in the cloud'' as the multi-armed bandit problem.
Finally, we discuss the heuristics to derive the exemplar configuration.

%\todo{we can extend to find multiple exemplar configurations}

\subsection{Empirical Study}
\label{sec:empirical}

We choose three popular software systems for cloud applications, namely Apache Hadoop 2.7, Spark 2.1 and Spark 1.5.
This study includes 30 applications for diversification.
They are data processing, OLAP queries, common statistics functions, and
popular machine learning algorithms.
Although they do not cover all the spectrum of real-world applications,
they are representative of many nowadays cloud applications.
When the input to applications changes, the workload behavior changes accordingly~\cite{Venkataraman2016, Dalibard2017}.
We also choose three different input parameters and data sizes for each application.
In total, our evaluation includes 107 workloads.

We conduct our evaluation on AWS EC2~\cite{aws}.
Regarding the VM to run the workloads, we choose 18 different VM types.
They include three instance families:
1) compute-optimized instances (\emph{c3} and \emph{c4}),
2) memory-optimized instances (\emph{r3} and \emph{r4}), and
3) general-purpose instances (\emph{m3} and \emph{m4}).
For each instance family, we choose \emph{large}, \emph{xlarge} and \emph{2xlarge} for the instance size.
Although we only evaluate 21\% VM types (AWS supports 85 kinds as of January in 2018),
they reflect many use cases on AWS EC2.
Besides, some VM types are designed for acceleration using GPU and FPGA, and therefore, they are less common and not included.
Furthermore, it is reported that VM types with lower than 8 cores dominate VM useage on Azure~\cite{Cortez2017}.
We try our best to reflect the common cloud deployment.
More details regrading data collection can be found in our previous work~\cite{Hsu2017, Hsu2018scout}.
We also made our data public available for further research~\cite{scoutdataset}.

\subsection{The Exemplar Configurations}
\label{sec:exemplar}
The exemplar configurations are configurations that
are near-optimal or satisfactory in the majority of workloads.
When the percentage is large, 
we can exploit the exemplar configurations to simplify collective optimization.
In \myfigure{\ref{fig:motivation_percentage}}, we present the opportunity of exploiting such configurations.
We count the number of the normalized performance that is within 30\% of the optimal.
The colored bars are possible exemplars because they are satisfactory at least in half of the workloads.
The red bar represents the VM type that is more likely to be the optimal than other configurations.
This figures show there exist several exemplar configurations.

% datasize=medium
\begin{table}[t]
\centering
\caption{\textbf{Normalized performance on a selected group of VM types and workloads.}  The number $1.0$ represents the optimal choice across the 18 VM types for the particular workload.}
\label{table:dataset}
%{\scriptsize
\resizebox{0.9\columnwidth}{!}{%
\renewcommand{\baselinestretch}{0.5} 
\begin{tabular}{@{}c@{~}l@{~}r@{~}r@{~}r@{~}r@{~}r@{}}
\toprule
\multicolumn{1}{l}{\textbf{System}} & \textbf{Workload} & \textbf{c3.large} & \textbf{c4.large} & \textbf{c4.xlarge} & \textbf{m4.large} & \textbf{m4.xlarge} \\ \midrule
\multirow{6}{*}{\rotatebox[origin=c]{90}{\parbox[c]{1.5cm}{\centering Hadoop 2.7}}} & aggregation & 1.26 & 1.00 & 1.12 & 1.12 & 1.29 \\
 & join & 1.26 & 1.00 & 1.09 & 1.17 & 1.26 \\
 & scan & 1.16 & 1.00 & 1.70 & 1.15 & 1.89 \\
 & sort & 1.10 & 1.00 & 1.06 & 1.03 & 1.11 \\
 & terasort & 1.31 & 1.00 & 1.16 & 1.07 & 1.12 \\
 & pagerank & 1.24 & 1.03 & 1.16 & 1.05 & 1.00 \\ \midrule
\multirow{10}{*}{\rotatebox[origin=c]{90}{\parbox[c]{1cm}{\centering Spark 2.2}}} & join & 1.12 & 1.00 & 1.40 & 1.12 & 1.23 \\
 & scan & 1.13 & 1.00 & 1.48 & 1.03 & 1.59 \\
 & sort & 1.11 & 1.00 & 1.42 & 1.13 & 1.40 \\
 & terasort & 1.30 & 1.19 & 1.66 & 1.34 & 1.46 \\
 & wordcount & 1.83 & 1.64 & 1.23 & 1.00 & 1.08 \\
 & als & 1.00 & 1.67 & 3.19 & 1.46 & 2.72 \\
 & aggregation & 1.30 & 2.00 & 1.08 & 1.00 & 1.18 \\
 & pagerank & 2.33 & 2.12 & 1.00 & 1.31 & 2.10 \\
 & bayes & 3.15 & 3.57 & 1.00 & 1.60 & 1.61 \\
 & lr & 6.50 & 5.56 & 1.44 & 1.00 & 2.61 \\ \midrule
\multirow{19}{*}{\rotatebox[origin=c]{90}{\parbox[c]{1cm}{\centering Spark 1.5}}} & chi-feature & 1.19 & 1.00 & 1.32 & 1.29 & 1.53 \\
 & fp-growth & 1.27 & 1.00 & 1.37 & 1.20 & 1.46 \\
 & gmm & 1.19 & 1.00 & 1.27 & 1.25 & 1.36 \\
 & gb-tree & 1.19 & 1.00 & 1.63 & 1.17 & 1.94 \\
 & pca & 1.16 & 1.00 & 1.11 & 1.15 & 1.31 \\
 & pearson & 1.19 & 1.00 & 1.11 & 1.19 & 1.11 \\
 & word2vec & 1.22 & 1.00 & 1.06 & 1.15 & 1.24 \\
 & spearman & 1.21 & 1.00 & 1.12 & 1.06 & 1.02 \\
 & statistics & 1.15 & 1.00 & 1.43 & 1.08 & 1.56 \\
 & svd & 1.16 & 1.00 & 1.02 & 1.07 & 1.09 \\
 & chi-gof & 1.24 & 1.12 & 1.46 & 1.00 & 1.81 \\
 & bayes & 1.27 & 1.15 & 1.19 & 1.25 & 1.35 \\
 & lda & 1.66 & 1.36 & 1.10 & 1.00 & 1.31 \\
 & pic & 1.53 & 1.39 & 1.00 & 1.15 & 1.31 \\
 & d-tree & 1.70 & 1.70 & 1.23 & 1.00 & 1.48 \\
 & als & 2.23 & 1.86 & 2.89 & 1.00 & 1.27 \\
 & regression & 4.03 & 3.60 & 4.06 & 4.42 & 4.70 \\
 & classification & 6.11 & 5.41 & 5.70 & 6.07 & 1.00 \\
 & kmeans & 6.22 & 5.74 & 3.66 & 3.73 & 1.00 \\ \midrule
\multicolumn{2}{c}{\textbf{\# of optimal}} & 1 & 18 & 3 & 7 & 3 \\
\multicolumn{2}{c}{\textbf{Mean}} & 1.89 & 1.72 & 1.63 & \textbf{1.45} & 1.53 \\
\multicolumn{2}{c}{\textbf{25\%}} & 1.18 & 1.00 & 1.11 & 1.04 & 1.15 \\
\multicolumn{2}{c}{\textbf{Median}} & 1.26 & 1.00 & 1.23 & 1.15 & 1.31 \\
\multicolumn{2}{c}{\textbf{75\%}} & 1.68 & 1.69 & 1.47 & \textbf{1.25} & 1.58 \\ \bottomrule
\end{tabular}
}
\end{table}

In \mytable{\ref{table:dataset}}, we give snippets of measurement data to better illustrate the exemplar configurations.
This table presents the normalized performance of workloads on some of the VM types.
A $1.0$ number indicates the VM type is the optimal choice for the corresponding workload while a larger number implies a sub-optimal choice.
We can observe that \emph{c4.large} is the optimal configuration for 18 workloads (out of 35).
However, it is also a sub-optimal VM type ($>1.4$) in 11 workloads, which generating
$1.72$ normalized performance on average.
On the other hand, \emph{m4.large} seems to be a better choice because it delivers $1.45$ performance on average and creates only 5 sub-optimal workloads.
The above gives one way to select the exemplar configuration and in the following,
we describe the challenges of selecting the exemplar.

\textbf{Varying workloads.}
In \mytable{\ref{table:dataset}}, we show five possible exemplar configurations for those particular workloads.
The exemplars very in different sets of workloads.
For example, \emph{c4.large} is the best choice in Hadoop 2.7 while
\emph{m4.large} should be selected as the exemplar VM type in Spark 2.2.

\textbf{Expanding cloud portfolio.}
As mentioned before, cloud providers introduce new VM types regularly, which
includes performance boost and price adjustment.
The exemplar configurations might also change accordingly.

\textbf{Online discovery.}
We present an offline analysis of measurement data above.
However, finding the exemplar configuration is an online task (for unknown workloads), which is considered a difficult learning problem.
This is similar to the exploration-exploitation dilemma~\cite{kaelbling1996reinforcement}.

From the above, it would appear that there exist exemplar configurations in real-world workloads.
Note that if the exemplar configuration is prevalent, it should be possible to simplify collective optimization as follows:
finding the exemplar configuration instead of finding the optimal choice for each of the workload.
The exemplar configurations deliver near-optimal to satisfactory performance in the majority of workloads.
The rest of this paper is a test of that speculation.

\subsection{Problem Formulation}
\label{sec:formulation}

\micky attempts to find the exemplar VM type ($vm^*\in VM$) for a group of workloads ($W$).
The workload refers to a combination of an application and the data used.
The performance is measured in terms of \textit{execution time} and \textit{operational cost}. The cloud configuration space for workload
$w$ is referred to as ($s\in S_w$), where $S_w$ is the set of cloud configuration options for a workload $w$. The size
of the search space is $N_w$ cloud configurations. In our setting, the size of the cloud configuration space is same for all workloads. For a given workload $w$, each configuration $s$ has a corresponding performance
measure $y_{w, s} = \phi(w, s)$.
%Each configuration in the cloud configuration space $s$ is represented as the features of the cloud configuration such as number of cores, size of the memory.
%The objective is to find a cloud

Single-optimizers such as Cherrypick~\cite{Alipourfard2017} searches a suitable VM type for every workload $w$ separately.
The search starts with a pool of unevaluated configuration ($U_w$)---the specific workload has not been run on any configuration.
As the search proceeds, the cloud configuration are selected from $U_w$ and moved to the evaluated pool ($E_w$). The sum of the cardinalities
of $U_w$ and $E_w$ is equal to the cardinality of $S_w$ ($|U_w| +|E_w| = |S_w|$).
The measurement cost of the search process is $C_w=|E_w|$.
When optimizing a group of workloads,
single-optimizers generate a total cost $C=\sum_{w \in W} |C_w|$.

\micky is a collective optimization method.
We explore the exemplar VM type $vm^*$ so that 
$|E_{w_1} \cup E_{w_2} \cup \cdots \cup E_{w_n}|$ is minimized while the corresponding performance measure
$y_{w, vm^*}$ is comparable to the the ones in single-optimizers.

\subsection{The Multi-Armed Bandit Problem}

To realize collective optimization, we reformulate the problem of configuration optimization as a multi-armed bandit problem~
\cite{robbins1985some,weber1992gittins,bergemann2006bandit,audibert2011introduction}.
In the problem setting,
an agent  (gambler) sequentially searches for a slot machine (from a group of slot machines) to maximize the total reward collected in the long run. This problem is non-trivial since the agent (gambler) cannot access the true probability of winning---all learning is carried out via the means of trial-and-error and value estimation. To find the suitable slot machine, the agent needs to acquire information about arms  (exploration)  while  simultaneously  optimizing  immediate rewards  (exploitation).
The is referred to as the exploration-exploitation dilemma~\cite{kaelbling1996reinforcement}.
Finding the better VM type for workloads naturally fits into the multi-armed bandit problem.
We describe their similarities in the following.

\textbf{Slot Machine.} Each VM type is similar to a slot machine. Our objective is to find the best VM that maximizes the reward for a group of workloads.

\textbf{Arm.} Arms are the choices of slot machines. In the cloud setting, an optimizer chooses a VM type to run a workload.

\textbf{Pull.} A pull is one play on the slot machine. It takes coins (cost) and yields a reward.
Similarly, an optimizer picks a VM type and measures the performance of a workload on the selected VM.

\textbf{Reward.} Reward refers to the amount of money a gambler wins or loses from pulling the arms.
In our setting, the reward is determined by where it meets a performance objective.
We use performance delta (between the selected and the optimal choice) for calculating the reward.
Please note that the optimal configuration is not known in the real-world setting.

\textbf{Budget.}
A gambler owns a certain amount to spend on the slot machines.
In our setting, an optimizer requires to complete the optimization process in a limited budget.
We use the number of measurements as the budget ($C$).
In practice, the minimal budget is usually $|\mathit{VM}|$
and the maximum budget is $|\mathit{VM}| \times |\mathit{W}|$.
The budget is determined by users.
A higher budget yields a better reward.

\textbf{Objective.}
The objective of \micky is to find the best configuration (minimize performance delta) for multiple workloads with fewer measurements.

The multi-armed bandit problems have attracted attention for solving online learning problems.
For example, 
Dambreville et al.~\cite{dambreville2017load} used multi-arm bandit to minimize the energy consumption of a cloud platform by
using workload prediction to reallocate the set of available servers.
Jiang et al. perform data-driven QoE (quality of experience) optimization for real-time exploration and exploitation~\cite{jiang2017pytheas}.
While we borrow techniques from this rich literature~\cite{bergemann2006bandit}, our contribution is to shed light on how to use these techniques to find the exemplar cloud configurations and to show collective optimization can solve the problem using only a fraction of measurement cost required by prior work.

\begin{figure*}[t]
 \centering
 \subfigure[Hadoop 2.7]{
 \label{fig:single_time_steps}
 \includegraphics[width=0.3\textwidth]{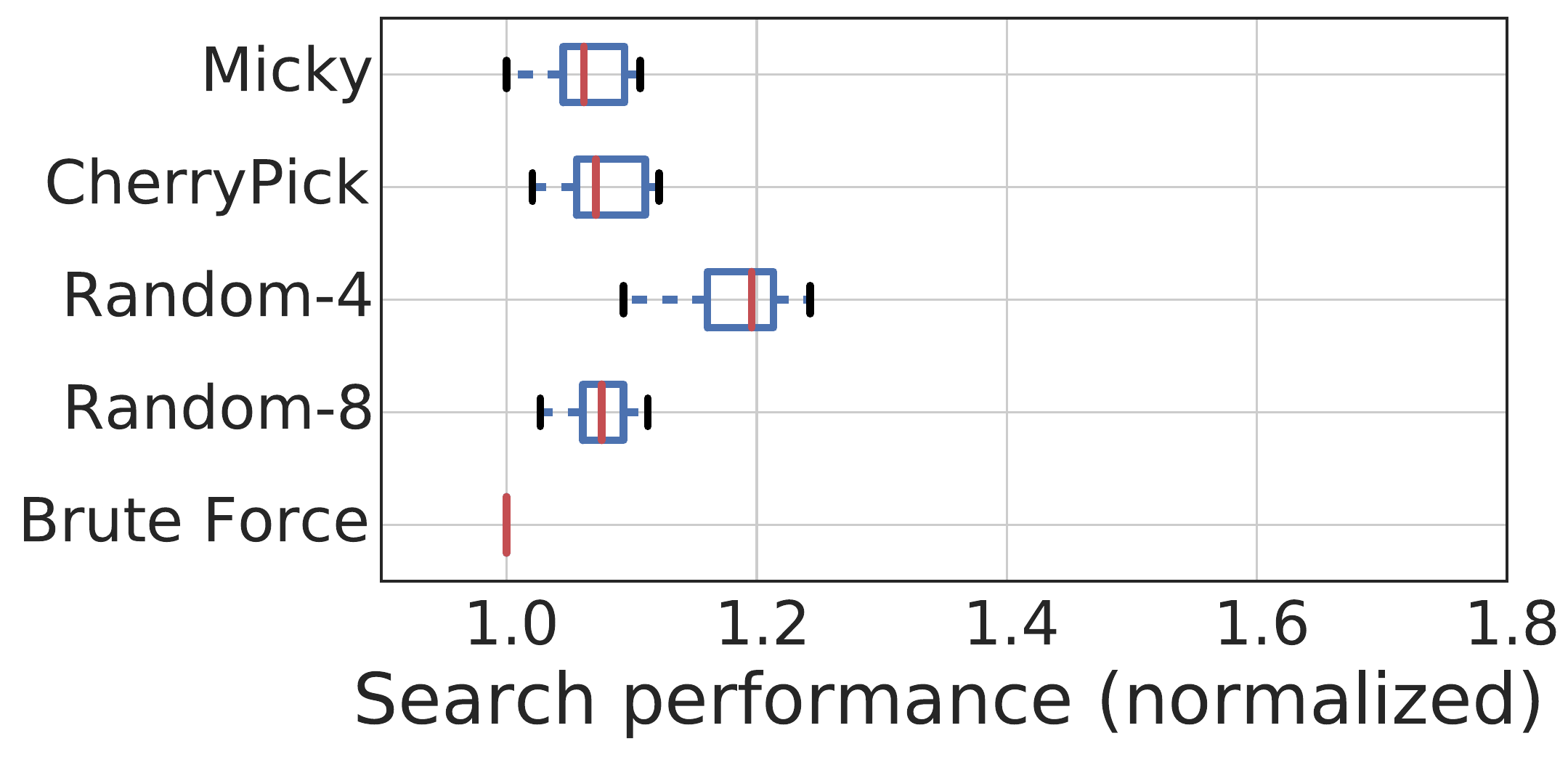}
 }
 \subfigure[Spark 2.1]{
 \label{fig:single_time_performance}
 \includegraphics[width=0.3\textwidth]{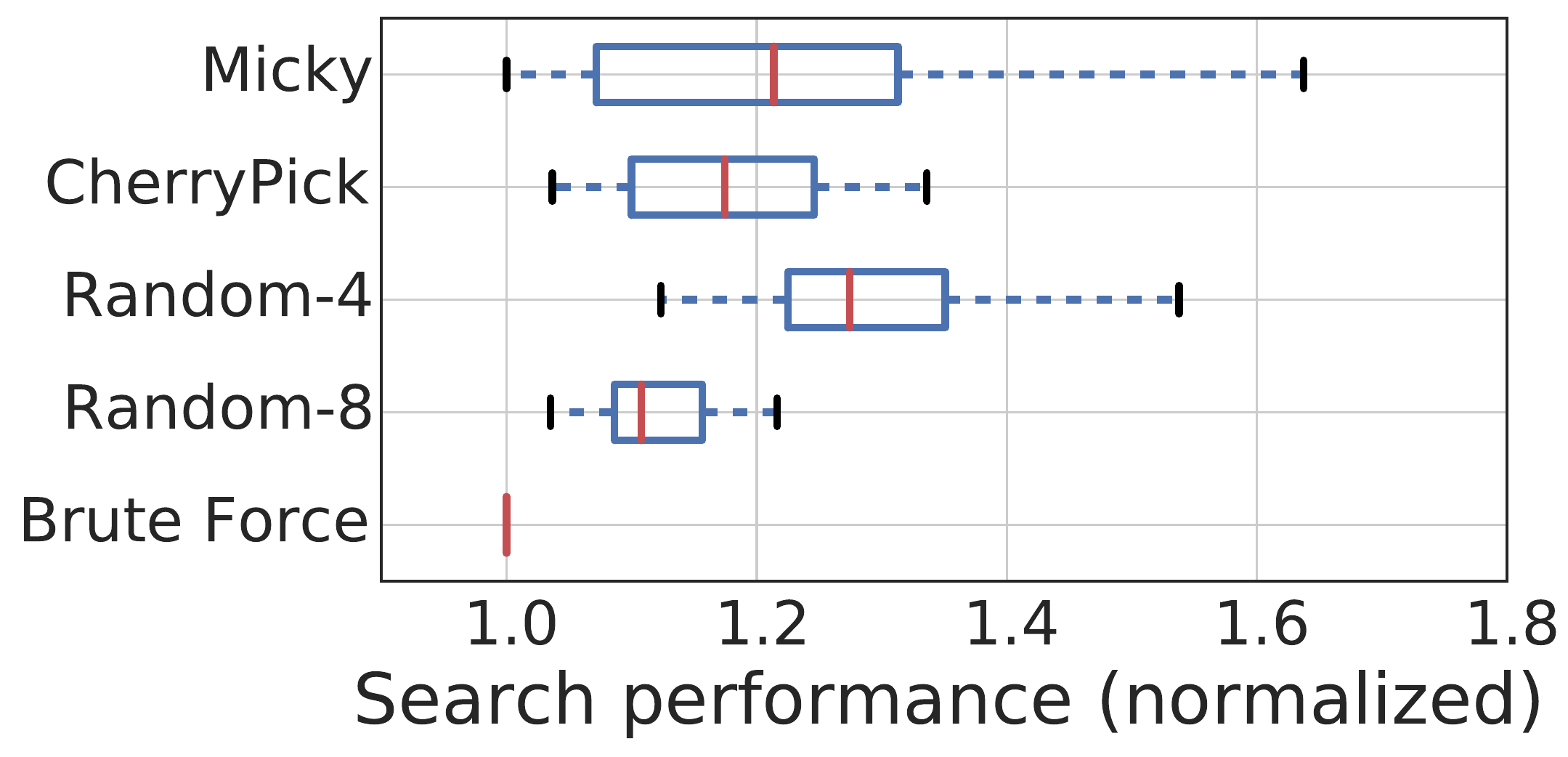}
 }
  \subfigure[Spark1.5]{
 \label{fig:single_time_performance}
 \includegraphics[width=0.3\textwidth]{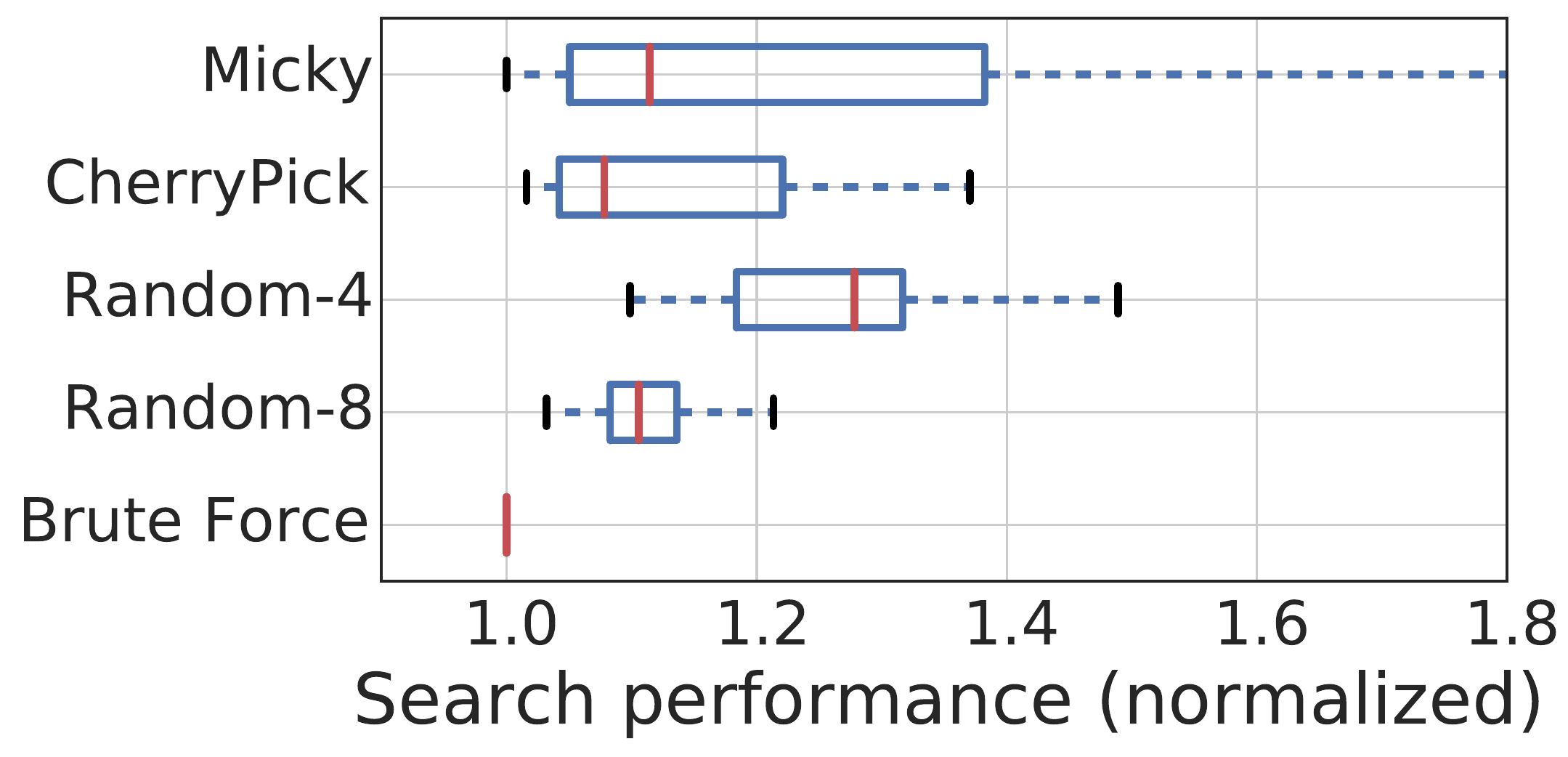}
 }
 \vspace*{-2mm}
 \centering
 \caption{\small{\textbf{Search performance of optimization methods in search for cost-effective cloud configurations.} Three software systems are evaluated.  \emph{CherryPick} finds good solutions in the three systems while \micky is comparable in Hadoop 2.7 but shows higher variance (sub-optimal choices).  We propose a integrated system (in \myfigure{\ref{fig:system_design}}) to detect those sub-optimal cases for improving \micky.}}
 \label{fig:s2_cost_performance}
 \vspace*{-4mm}
\end{figure*}

\subsection{Heuristics}
\label{sec:methods}

% need to put softmax in the following text
% we might need to try thompson sampling
% need citation
In the literature, several strategies have been proposed to find the most rewarding slot machines (the exemplar configurations) in the multi-arm bandit setting.
These strategies can be divided into three major groups.
First, the \textit{Epsilon-greedy}, works by oscillating between (a) exploiting the best option which is currently known, and (b) exploring at random among all of the options available to it.
Second, the probability matching strategy selects the arms according to the probability of the arm being the optimal choice.
Thompson sampling or Bayesian Bandits are well-known probability matching strategies.
Last, in the contextual bandit problem, strategies such as Upper Confidence Bound (UCB) builds a predictor from existing observation for making a better decision.
UCB always opportunistically chooses the arm that has the highest upper confidence bound of reward, and therefore, it will naturally tend to use arms with high
expected rewards or high uncertainty.
The above only discusses some important methods.
It is not the major focus to design the best method but to evaluate the existing methods best for collective optimization.

%%%%%%%%%%%%%%%%%%%%%%%%%%%%%%%%%%%%%%%%%%%%%%%%%%%%%%%%%%%%%%%%%%%%%%%%
%  Evaluation
%%%%%%%%%%%%%%%%%%%%%%%%%%%%%%%%%%%%%%%%%%%%%%%%%%%%%%%%%%%%%%%%%%%%%%%%

\section{Evaluation}
\label{sec:evaluation}
\vspace{-2mm}

%This section describes our experimental design and the evaluation results.

\subsection{Comparison Method}

We compare our method with \emph{Brute Force}---measures all possible configurations and \emph{CherryPick}---the state-of-the-art method
~\cite{Alipourfard2017}.
Please refer to the related work (Section~\ref{sec:related_work}) for more details.
Besides, we use  \emph{Random-4} and \emph{Random-8}, which randomly measures 4 and 8 configurations (for each workload) respectively as straw man methods.
The comparison metrics are  \emph{measurement cost} and \emph{search performance}.

The \emph{brute force} approach needs to test each configuration, and therefore, it generates constant measurement cost ($|S_w| \times |W|$).
The measurement cost of CherryPick varies for different workloads since it uses a heuristic stopping criterion. The lower bound is $3\times |W|$ because CherryPick uses at least three measurements as its initial points.
\micky performs collective optimization, and hence the measurement cost is shared by a batch of workloads and therefore, expected to be much lower than the other methods.

To compare their search performance, we use normalized performance in terms of execution time (the elapsed time required to complete a workload) and operational cost (the charge for completing a workload).
The \emph{brute force} approach always finds the optimal configuration while
the CherryPick is very likely to find near-optimal choices.
We examine whether \micky can find near-optimal configurations that are comparable to the \emph{CherryPick} approach.
A method delivers better search performance when the performance of found solutions is closer to the optimal.
For example, $1.05$ is better than $1.15$ because the former is only 5\% slower than the optimal.

We demonstrate the effectiveness of \micky using 107 workloads on Apache Hadoop and Spark.
These workloads are representative of many real-world applications.
Table~\ref{table:dataset} lists a subset of the workloads.
Please refer to our previous work for more details~\cite{Hsu2017, Hsu2018scout, scoutdataset}.

\iffalse
\begin{figure}[t]
 \includegraphics[width=.4\textwidth]{Figures/s0_single_time_steps_sum.pdf}
 \centering
 \caption{\textbf{Growth of measurement cost.} \micky is more economic when the number of workloads increases.  When the budget for optimization is limited, collective optimization becomes more promising.}
 \label{fig:scalability}
\end{figure}
\fi

\subsection{Experiment Setup}
\label{sec:setup}

%We list the parameters used in \emph{CherryPick} and \micky. 

{\textbf{CherryPick---Bayesian Optimization}}:
We encode the cloud configurations (\ie{CPU types, core counts, memory size per code and the bandwidth to Elastic Block Storage}) to represent the search space.
For the parameters, we choose the same kernel function (\emph{Mat\'ern 5/2}) and
the same stopping criteria (\textbf{EI}=10\%), as used in \emph{CherryPick}.
Regarding the choice of initial points, we randomly select three cloud configurations.
The above process is repeated 100 times for reducing artifact and better showing the capability of \emph{CherryPick}.

{{\sc \textbf{Micky}}\textbf{---Multi-Armed Bandit}}:
There are three common algorithms for the multi-armed bandit problems as described in Section~\ref{sec:mab}.
We choose UCB because it is more stable as compared to other bandit algorithms
(will be discussed later in Section~\ref{sec:tuning_mab}).
\micky runs in two phases: (1) pure exploration, and (2) exploration along with exploitation.
In the pure exploration phase, \micky measures the performance of VMs with random workloads for improving stability and reduces sampling bias.
The $\alpha$  parameter represents the number of exhaustive iterations over each VM type.
In the second phase, \micky runs the algorithm to handle the exploration and the exploitation.
The behavior of this phase is controlled by the parameter $\beta$, which controls the number of measurements for finding the exemplar configurations.
The measurement cost of \micky is $(\alpha \times \mathit{|S|} + \beta \times \mathit{|W|}$).
We have observed that the measurement cost is directly proportional to the effectiveness of \micky.
In our experiments, we choose $\alpha=1$ and $\beta=0.5$.

\subsection{Can \emph{Micky} identify the exemplar cloud configurations?}

The primary goal of \micky is to find the most suitable cloud configuration across all workloads.
In this evaluation, we show the search performance in finding the cost-effective VM types.
In Figure~\ref{fig:s2_cost_performance}, we use box plot for comparison.
The red line in the box represents the median value while the two sides of the box are the first and third quartile.
The whiskers represent the 10 and 90 percentile respectively.

From this figure, we observe that the performance of \micky is comparable to CherryPick in the majority of workloads (using the median).
\micky is only 5\% worse than CherryPick on Spark 2.1 and Spark 1.5.
Surprisingly, \micky is slightly better than CherryPick on Hadoop 2.7.
The variance of \micky is higher because \micky
optimizes most workloads but fails to optimize for some.
We will discuss how to remedy this situation in Section~\ref{sec:system}.

To explain why \micky works, we further analyze the exemplar VM types recommended by \micky as listed in \mytable{\ref{table:top3}}.
The table shows the percentage of workloads that are within the performance thresholds.
CherryPick finds good VM types ($< 1.2$) in 86\% of workloads while
\micky achieves the same search performance performance in 71\% of workloads using only 11.6\% of measurements by CherryPick.

% \usepackage{booktabs}
\begin{table}[t]
\centering
\caption{\textbf{The most cost-effective VM types  for 107 workloads recommended by \micky} The number above each column label represents normalized performance (to the optimal). CherryPick finds good ($< 1.2$) VM types in 86\% of workloads.}
\label{table:top3}
\begin{tabular}{@{}lp{8mm}p{8mm}p{8mm}p{10mm}p{10mm}@{}}
\toprule
& $= 1.0$ \newline \textbf{Optimal}   & $< 1.1$ \newline \textbf{Excellent} & $< 1.2$ \newline \textbf{Good} & $\le 1.4$ \newline \textbf{Tradeoff} & $> 1.4$ \newline \textbf{Unsettled}     \\ \midrule
c4.large  & 48\%      & 61\% & 66\%      & 70\%        & 30\% \\
m4.large  & 27\%      & 46\% & 71\%      & 84\%        & 16\% \\
m4.xlarge & 9\%       & 15\% & 32\%      & 63\%        & 37\% \\ \bottomrule
\end{tabular}
\end{table}

\subsection{When not to use \micky?}
\label{sec:kneepoint}

\micky reduces measurement cost while delivering satisfactory performance.
However, there exists a trade-off between the cost reduction achieved by \micky and its effectiveness.

\begin{figure}[t]
 \includegraphics[width=.4\textwidth]{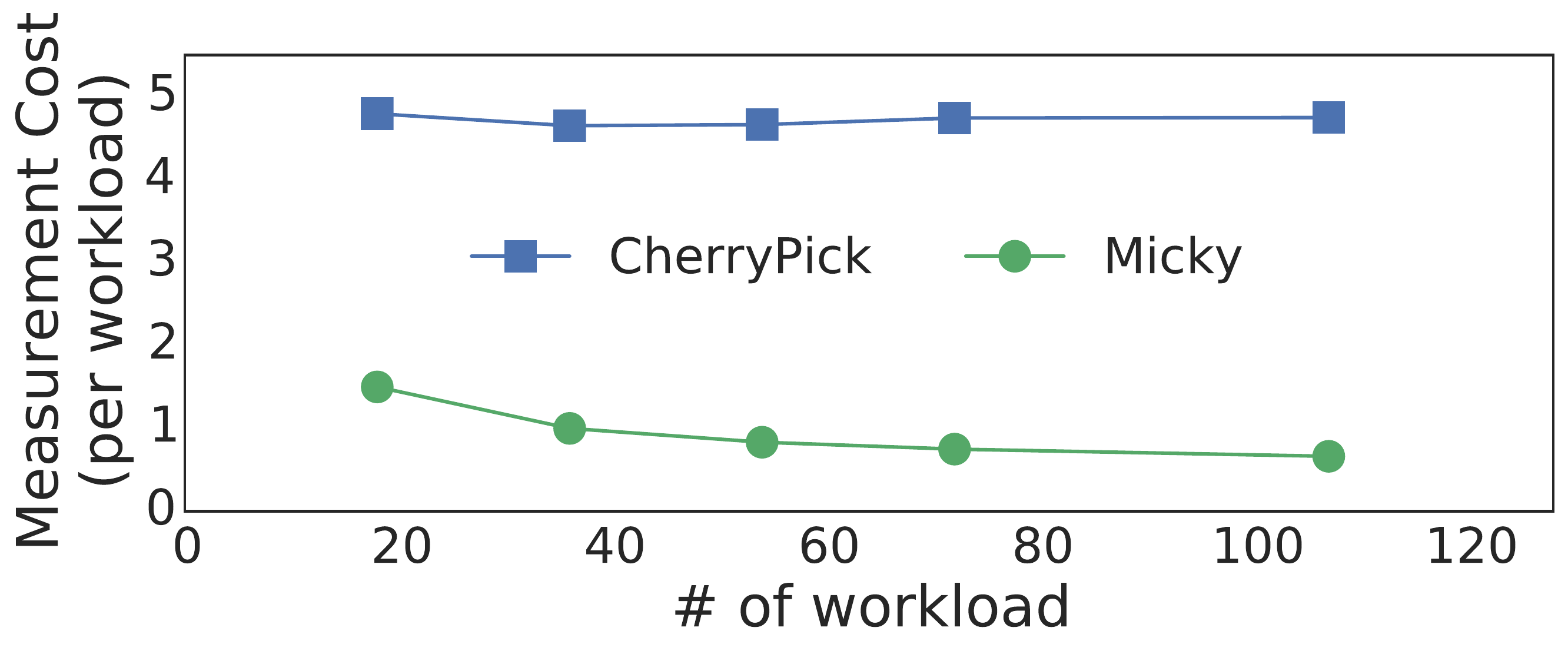}
 \vspace*{-2mm}
 \centering
 \caption{\textbf{Low measurement cost in collective optimization.} \emph{CherryPick} optimizes each workload separately while \micky finds the exemplar cloud configuration suitable for a group of workloads.}
 \label{fig:cost_saving}
 \vspace*{-4mm}
\end{figure}

First, \myfigure{\ref{fig:cost_saving}} shows that \emph{CherryPick} is four times expensive than \micky. As the number of workloads increases, \micky is more economical because
the cost of pure exploration phase of \micky remains constant. This is because $\alpha$ depends only on the number of cloud configurations.
We observe that \micky only uses a fraction of the measurement cost when compared to the other methods.
For example, when optimizing for 40 workloads, \micky only uses 30 measurements to find the suitable cloud configuration whereas CherryPick uses 156 measurements.
Another observation is that \micky is more scalable because the slope of the line decreases as the number of workloads increases.

Second, a user demands near-optimal solutions (\eg{$<1.1$})
mostly for highly recurring workloads because its measurement cost can be amortized.
In \mytable{\ref{table:break_even_cost}}, we show the \textit{knee-point} that a user should use a single-optimizer rather thatn a collective-optimizer.
We calculate the knee point using
$K \times f(\Delta_{P}, C_{P}) \ge g(\Delta_{M}, C_{M})$,
where $K$ is the recurrence of a workload as the knee point,
the function $f$ represents the opportunity loss due to inferior search performance, and
the function $g$ represents the reduction of measurement cost when using collective optimization.
In addition, 
$\Delta_{P}$ is the delta of normalized search performance (between a single- and collective-optimizer),
$\Delta_{M}$ is the delta of measurement cost.
$C_{P}$ and $C_{M}$ are cost (\eg{dollars}) defined by users.
For simplification, we use $C_{P} = 10 \times C_{M}$ in this calculation..
For non-critical workloads (\eg{recurring batch-process jobs}),
$C_{P}$ is lower, and hence, \micky is more beneficial.
As shown in \mytable{\ref{table:break_even_cost}},
\emph{CherryPick} is preferred only when
the same workloads run more than 20 to 30 times.
Otherwise, \micky is a more desirable solution.

% \usepackage{booktabs}
\begin{table}[t]
\centering
\caption{\textbf{The knee point when \micky should not be used.} 
The knee point (the number of recurrence of workloads) represents a trade-off between search performance and measurement cost.
}
\label{table:break_even_cost}
\begin{tabular}{@{}llllll@{}}
\toprule
            & 18   & 36    & 54   & 72   & 107  \\ \midrule
Brute Force & 84.8 & 120.6 & 55.0 & 52.1 & 57.3 \\
Random-8    & 37.9 & 51.5  & 33.7 & 36.0 & 44.7 \\
Random-4    & 18.4 & 24.2  & 27.0 & 28.5 & 27.9 \\
CherryPick  & 23.3 & 30.8  & 20.8 & 24.0 & 27.0 \\ \bottomrule
\end{tabular}
\end{table}

\begin{figure}[t]
 \includegraphics[width=.4\textwidth]{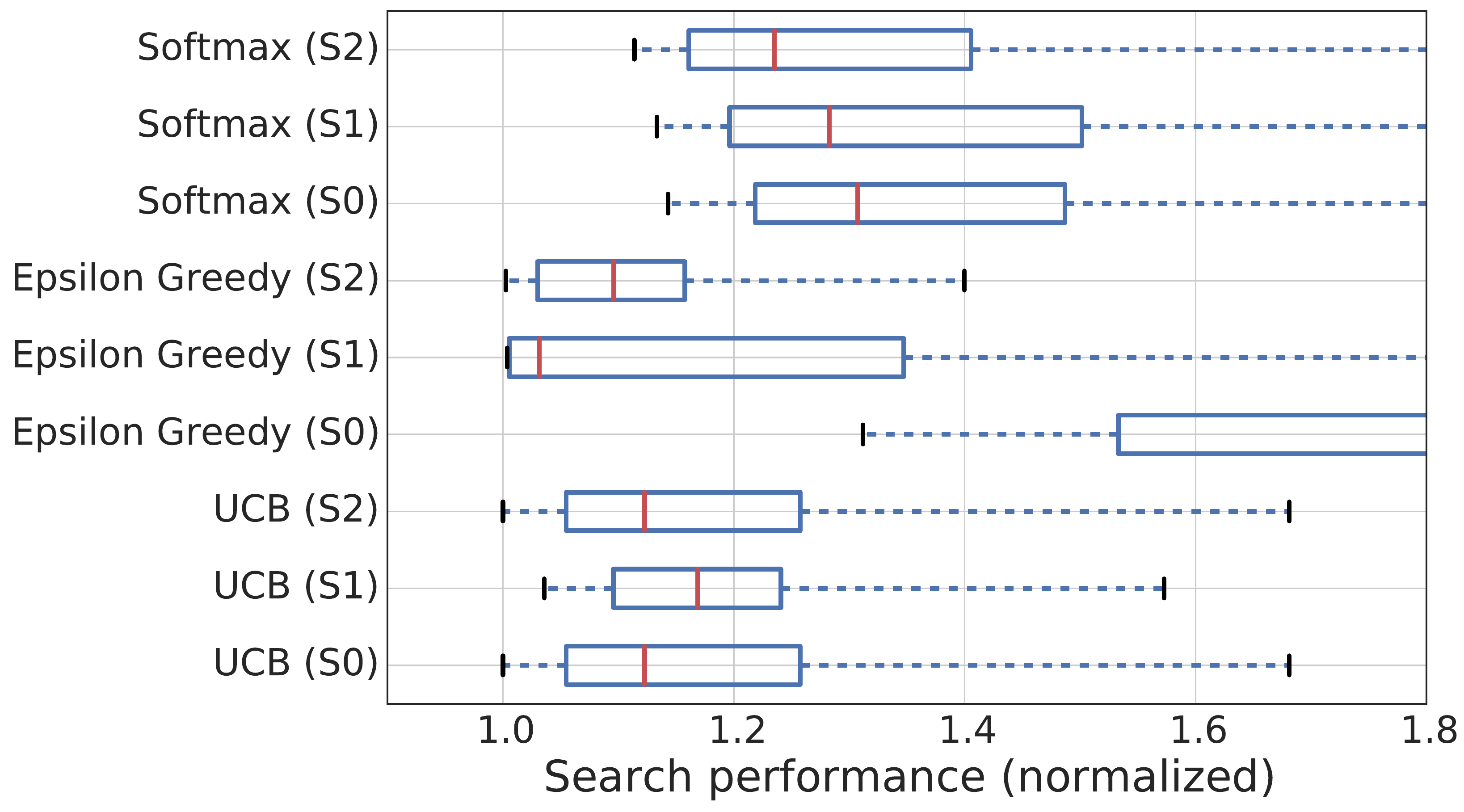}
 \vspace*{-2mm}
 \centering
 \caption{\textbf{Selection of multi-armed bandit algorithms.} The parameter (in the parenthesis) controls the measurement budget ($S0 < S1 < S2$).}
 \label{fig:algorithm_selection}
 \vspace*{-4mm}
\end{figure}

\subsection{Why UCB is the preferred choice?}
\label{sec:tuning_mab}

To select the suitable method for \micky, we compare three multi-armed bandit algorithms
(as mentioned in Section~\ref{sec:methods}).
First, the behavior of Epsilon Greedy is controlled by the parameter $\epsilon$. A larger value encourages exploration while a lower value encourages exploitation.
We choose 0.1 for the epsilon parameter.
Second, the Softmax algorithm uses a temperature parameter for structured exploration.
The Softmax algorithm with an infinity temperature uses pure exploration while a zero value sticks to the arm (cloud configuration) with the highest estimated probability---pure exploitation.
We use 0.1 for the temperature parameter.
Last, the Upper Confidence Bound algorithm (UCB) tracks the confidence of rewards of arms.
There are no parameters.

\myfigure{\ref{fig:algorithm_selection}} presents the comparison between the three methods.
The parameter in the parenthesis represents the measurement budget, determined by $\alpha$ and $\beta$ (as described in Section~\ref{sec:setup}).
We choose 0, 1, 2 as the $\alpha$ parameter for $S0, S1, S2$ and use 0.5 for $\beta$ in all.
This figure shows UCB is more stable.
Besides, the performance of UCB does not heavily rely on parameter tuning.
Therefore, we prefer UCB to Epsilon Greedy, and \micky is built using UCB.

\begin{figure}[t]
 \includegraphics[width=.45\textwidth]{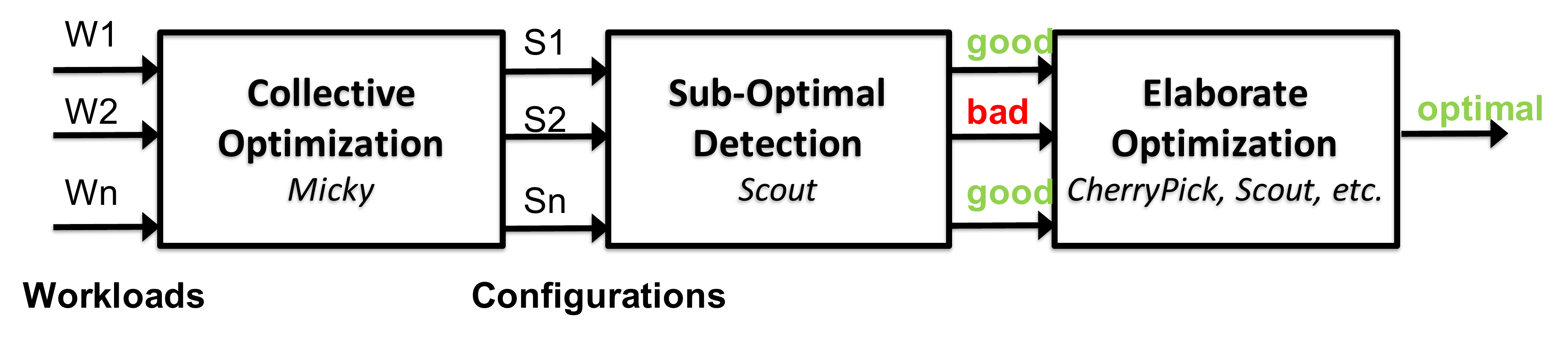}
 \vspace*{-2mm}
 \centering
 \caption{\textbf{A system integration to alleviate sub-optimal choices in some workloads.} \scout answers ``is there a better configuration than the current choice?''~\cite{Hsu2018scout}. An integration of \micky and \scout delivers a more efficient and reliable recommendation system of cloud configurations.}
 \label{fig:system_design}
 \vspace*{-4mm}
\end{figure}

%%%%%%%%%%%%%%%%%%%%%%%%%%%%%%%%%%%%%%%%%%%%%%%%%%%%%%%%%%%%%%%%%%%%%%%%
%  System Design
%%%%%%%%%%%%%%%%%%%%%%%%%%%%%%%%%%%%%%%%%%%%%%%%%%%%%%%%%%%%%%%%%%%%%%%%
\section{To Eliminate Sub-Optimal Choices}
\label{sec:system}

While the exemplar configuration is adequate for most workloads and reduces measurement cost significantly, it is almost inevitably that fewer workloads may suffer from sub-optimal performance (since we trade-off near-optimal performance for a large reduction in measurement cost).
For instance, 30\% workloads (\textit{c4.large}) underperform ($> 1.4$) as shown in \mytable{\ref{table:top3}}.
Similarly, the 90 percentile in \myfigure{\ref{fig:s2_cost_performance}}.
Although we have shown \micky is much more practical in the knee point analysis (Section~\ref{sec:kneepoint}),
it would be great if we can inform users of those sub-optimal choices.

\begin{figure}[t]
 \includegraphics[width=.4\textwidth]{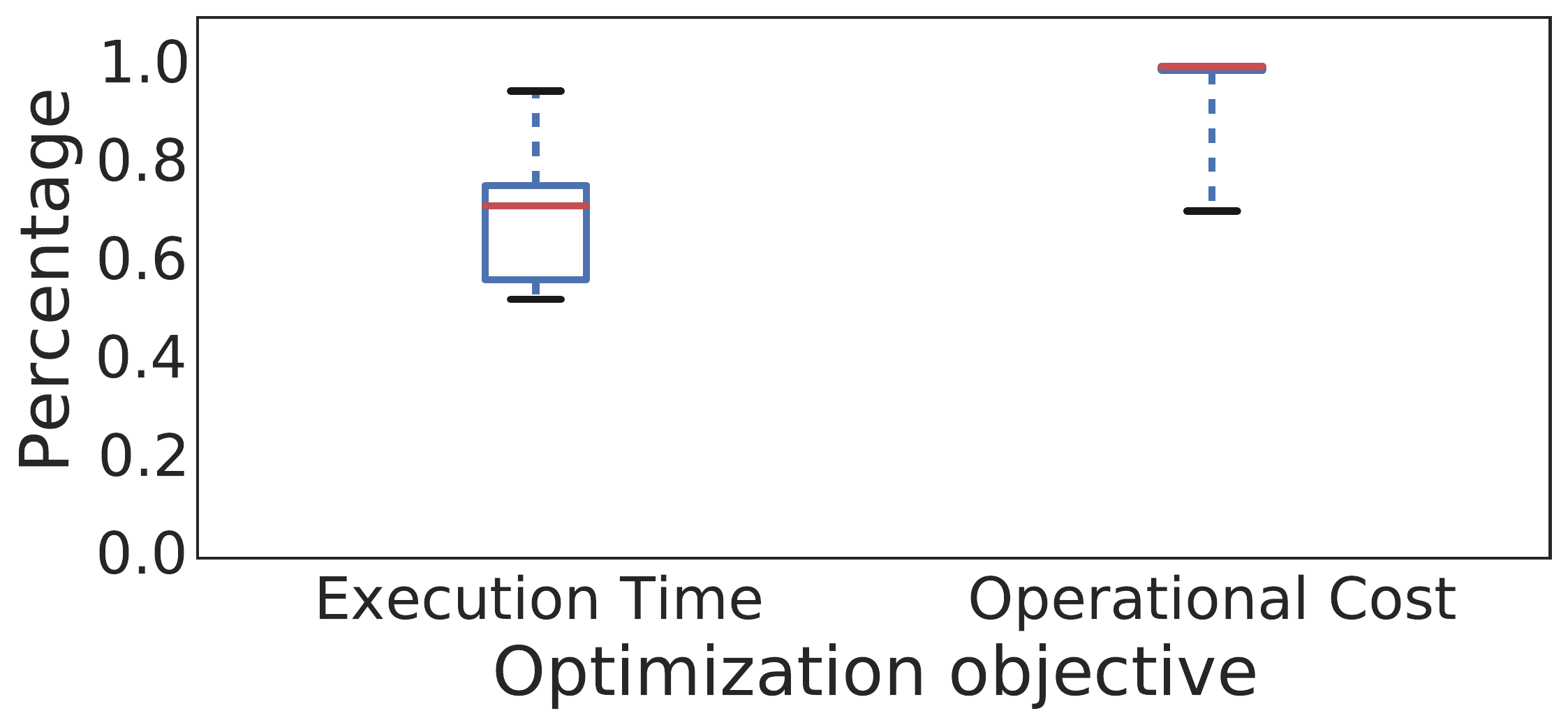}
 \vspace*{-2mm}
 \centering
 \caption{\textbf{Detection of mis-predictions using \scout.} The percentage represents the truth positive ratio, the probability the unsettled configurations can be identified.  The two optimization objectives are to find the fast configuration and the most cost-effective VM type respectively.}
 \label{fig:detection_misprediction}
 \vspace*{-4mm}
\end{figure}

We propose a two-level approach that integrates our previously built system, \scout, to detect this problem for further optimization~\cite{Hsu2018scout}.
\scout is able to answer ``is there a better configuration than the current choice?''.
\myfigure{\ref{fig:system_design}} illustrates the proposed system integration.
Users get choices of optimizing those under-performed workloads.
\myfigure{\ref{fig:detection_misprediction}} indicates that those sub-optimal choices are very likely to be identified.
The detection module can detect bad performance with a median accuracy of 98\%.
This is promising because users benefit from
low measurement cost (by \micky) and
performance guarantee (by \scout).
This ability enables users to further optimize for those sub-optimal workloads, which is particularly beneficial to highly recurring workloads.

%%%%%%%%%%%%%%%%%%%%%%%%%%%%%%%%%%%%%%%%%%%%%%%%%%%%%%%%%%%%%%%%%%%%%%%%
%  Related Work
%%%%%%%%%%%%%%%%%%%%%%%%%%%%%%%%%%%%%%%%%%%%%%%%%%%%%%%%%%%%%%%%%%%%%%%%

\section{Related Work}
\label{sec:related_work}

%This section describes several important cloud computing optimizers.
%We discuss their trade-offs and constraints, and then
%we show why we need a cloud optimizer for multiple workloads.
%In the end, we provide a guild to the selection of cloud optimizers in practice.

%\subsection{Cloud Computing Optimizer}

The cloud computing optimizer determines the best cloud configuration (such as VM types and cluster sizes) for a given workload.
Users are looking for configurations that are
highly performing (\eg{the shortest execution time}) or
cost-effective (\eg{the cheapest operational cost}),
or meeting the trade-off between them~\cite{Alipourfard2017,Yadwadkar2017,Hsu2017}.
A poor choice, for example, can lead to
a 20 time slowdown or a 10 times increase in total cost~\cite{Hsu2017}.
Although cloud providers recommend the choice of VM types,
it is too coarse grain to be effective~\cite{aws, google_rightsizing}.
Besides, resource requirement for meeting a certain objective
is opaque~\cite{Yadwadkar2017}.
Previous attempts are listed as follows.

\textbf{Ernest}
exploits the internal structure of the workload to predict execution time of a workload~\cite{Venkataraman2016}. This significantly reduces measurement cost. However, \emph{Ernest} is not scalable because the prediction model is specific to a VM type.

\textbf{PARIS} uses historical data to build a learning model for predicting performance and cost of workloads on different VM types~\cite{Yadwadkar2017}.
Building an accurate model requires comprehensive training data to cover diverse workload characteristics.
Besides, it may suffer from high prediction error (as high as 50\%) in batch-processing workloads~\cite{Yadwadkar2017}. 

\textbf{CherryPick} uses Bayesian Optimization, which updates its beliefs (workload performance on configurations) and finds the best configuration sequentially~\cite{Alipourfard2017}.
Although Bayesian Optimization is powerful, it can be fragile when the search space is not well represented~\cite{Hsu2017}.

\textbf{Arrow} leverages low-level performance metrics to address the fragility issue in \textbf{CherryPick} due to insufficient representation in the search space and poor choices of the kernel function in Gaussian Process~\cite{Hsu2017}.

\textbf{Scout} uses historical data and leverages low-level performance metrics~\cite{Hsu2018scout}.
This approach improves model accuracy, solves the cold-start issue and alleviates the fragility issue.

In the literature, software configuration optimization~\cite{herodotou2011starfish,zhu2017bestconfig,bilal2017towards,Dalibard2017}, program parameter tuning~\cite{Klein2017,golovin2017google} and sampling techniques~\cite{nair2018finding, oh2017finding, Nair2017} are active research directions.
They all focus on the same machine configuration.
It is not clear how to apply their approach directly to cloud environments, where
workloads perform very differently on distinct cloud configurations, \eg{VM types}.

%%%%%%%%%%%%%%%%%%%%%%%%%%%%%%%%%%%%%%%%%%%%%%%%%%%%%%%%%%%%%%%%%%%%%%%%
%  Guide
%%%%%%%%%%%%%%%%%%%%%%%%%%%%%%%%%%%%%%%%%%%%%%%%%%%%%%%%%%%%%%%%%%%%%%%%

\section{A Practical Guide to Cloud Optimizer}
\label{sec:guide}

%\subsection{Selecting the Right Optimizer}

\begin{figure*}[t]
 \includegraphics[width=.75\textwidth]{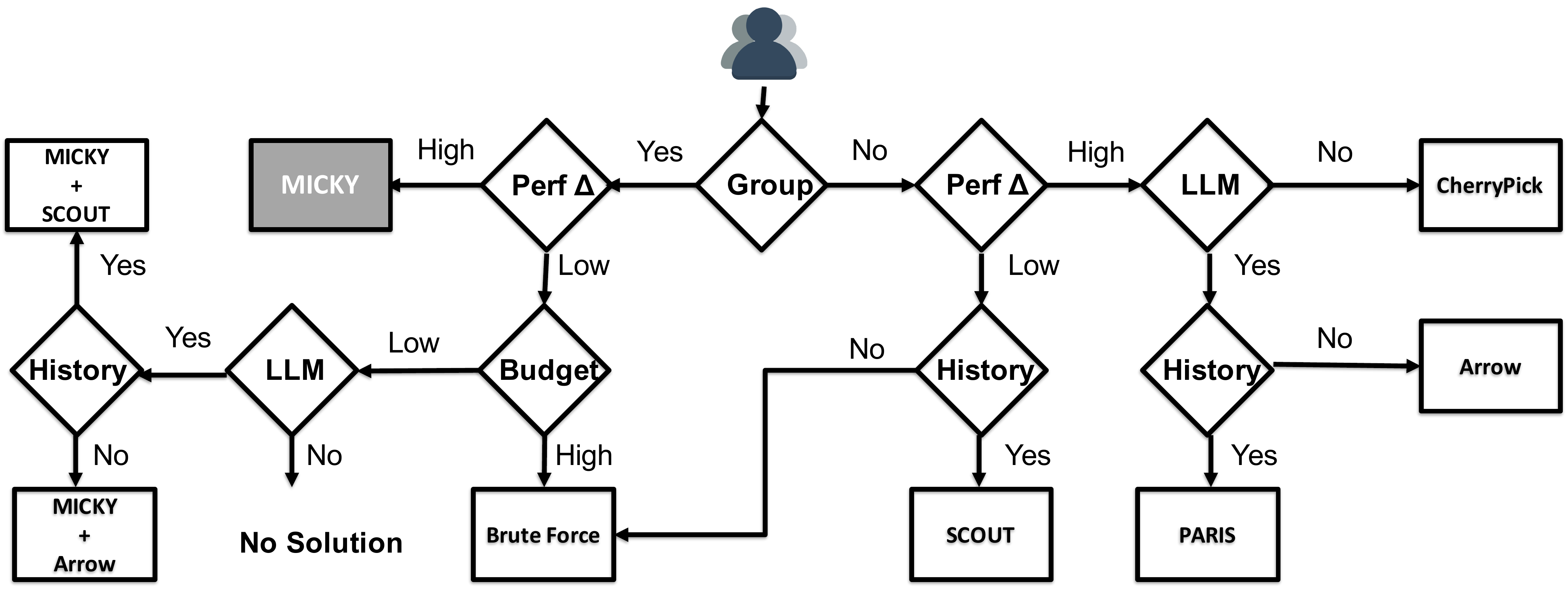}
 \vspace*{-2mm}
 \centering
 \caption{\textbf{A practical guide to choosing the right optimization method.} \emph{CheeryPick} works for any workloads without historical and low-level performance data~\cite{Alipourfard2017}. \emph{Arrow} uses low-level metrics to augment Bayesian Optimization (used in \emph{CherryPick})~\cite{Hsu2017}.  \emph{PARIS} requires low-level and historical data for predicting execution time and running cost of workloads on different VM types~\cite{Yadwadkar2017}.  \scout leverages a learning model and sequential model-based optimization (SMBO) to deliver efficient, effective and reliable recommendation~\cite{Hsu2018scout}.  \emph{Micky}, different from others, applies collective optimization to largely reduce measurement cost.}
 \label{fig:practical_guide}
 \vspace*{-4mm}
\end{figure*}

To pick an optimizer, we should compare its search performance and measurement cost, and understand their assumptions and constraints.
In \myfigure{\ref{fig:practical_guide}}, we derive a practical guide
for selecting an optimizer.
This guide is derived based on extended literature review and our extensive experimentation.

\textbf{Performance delta (Perf $\Delta$)}
represents search performance, the lower, the better.
Some cloud optimizers may suffer from
the fragility issue or high prediction error.
They are considered less reliable.
When using these optimizers, users should be more careful
because they do not know whether the recommended configurations by the optimizers are near-optimal or sub-optimal.

\textbf{Low-level Metrics (LLM)} are runtime information (such as CPU utilization, memory usage, and I/O rates) for better characterizing
workloads.
If such low-level information is accessible, users should choose optimizers that leverage low-level performance information.

\textbf{Historical data (History)} is execution records of workloads on cloud configurations.
\emph{CherryPick} and \emph{Arrow} do not use historical data (from other workloads) and therefore, require significant initial measurements for 
building prediction models while
\emph{PARIS} and \emph{Scout} uses historical data.

\textbf{Budget} is the measurement cost a user is willing to pay for an optimizer.
While the brute force approach delivers the best search performance,
it is too expensive in practice.
Using the state-of-the-art methods, for example,
\emph{CherryPick} incurs measurement cost of about 22\% to 33\% of the configuration space, and
\emph{Scout} reduces the cost down to 11\% to 19\% while achieving similar or better search performance~\cite{Hsu2018scout}.

\myfigure{\ref{fig:practical_guide}} summarizes the contribution of this paper.
\micky reduces measurement cost while delivering comparable search performance for a group of workloads.
To address the sub-optimal choices in some workloads, we propose an integration with \scout for further optimization.

%%%%%%%%%%%%%%%%%%%%%%%%%%%%%%%%%%%%%%%%%%%%%%%%%%%%%%%%%%%%%%%%%%%%%%%%
%  Limitation
%%%%%%%%%%%%%%%%%%%%%%%%%%%%%%%%%%%%%%%%%%%%%%%%%%%%%%%%%%%%%%%%%%%%%%%%

\iffalse
\section{Threats to Validity}
\label{sec:TOV}

\vspace{0.16cm}
\noindent {\textbf{Sampling bias:}} Our empirical study and evaluation are based on Apache Hadoop and Spark with a limited number of workloads.
It is still not clear whether this phenomenon is prevalent in other systems, applications, and configurations.

\noindent \textbf{Internal Validity:} The parameter setting can be considered a threat since an ideal tuning of parameters was not performed by time constraints. To minimize this threat, we used the parameters defined in related work. 

\noindent \textbf{Evaluation bias}: This paper measures performance using search performance and cost. There could be other ways to measure the performance such as resource requirements, the actual cost of evaluation, etc.

\noindent \textbf{Conclusion Validity:}
Since many factors affect search performance,
for a fair comparison,
we repeat our experiments 100 times and report the median and the variance of the results.
\fi

%%%%%%%%%%%%%%%%%%%%%%%%%%%%%%%%%%%%%%%%%%%%%%%%%%%%%%%%%%%%%%%%%%%%%%%%
%  Conclusion
%%%%%%%%%%%%%%%%%%%%%%%%%%%%%%%%%%%%%%%%%%%%%%%%%%%%%%%%%%%%%%%%%%%%%%%%

\section{Conclusion}
\label{sec:conclusion}
Collective optimization is promising and yet practical for deploying multiple workloads in clouds.
The collective optimization problem is similar to the multi-armed bandit problem.
With existing heuristics, we are able to derive the exemplar cloud configuration that works well across a group of workloads. Collective optimization greatly reduces measurement cost while producing optimal to satisfactory performance.

%\section*{Acknowledgment}



%\bibliographystyle{IEEEtran}
%\begin{thebibliography}{1}


%\end{thebibliography}

\bibliographystyle{IEEEtran}

%\bibliography{paper}

\end{document}